\documentclass[prc,aps,a4paper,groupedaddress,superscriptaddress,nofootinbib
,preprintnumbers,twocolumn]{revtex4}
\usepackage{graphicx}
\usepackage{amsfonts}
\usepackage{amssymb}
\usepackage{amsmath}
\usepackage{natbib}
\usepackage{dcolumn}
\usepackage{bm} 
\usepackage{color,xcolor}
\usepackage{multirow}
\usepackage{caption}
\usepackage{hyperref}
\usepackage[normalem]{ulem}
\usepackage[export]{adjustbox}
\usepackage{float}
\usepackage{scrextend}

\newcommand{\bwt}{\begin{widetext}}
	\newcommand{\ewt}{\end{widetext}}
\newcommand{\beq}{\begin{equation}}
	\newcommand{\eeq}{\end{equation}}
\newcommand{\bea}{\begin{eqnarray}}
	\newcommand{\eea}{\end{eqnarray}}

\begin{document}

\setlength{\tabcolsep}{10pt}
\title{Assessing the joint effect  of temperature and magnetic field on the neutron star equation of state}
\author{Luigi Scurto}
\email{lscurto@student.uc.pt}
\affiliation{CFisUC, Department of Physics, University of Coimbra, 3004-516 Coimbra, Portugal}
\author{Val\'eria Carvalho} 
\email{val.mar.dinis@uc.pt} 
\affiliation{CFisUC, Department of Physics, University of Coimbra, 3004-516 Coimbra, Portugal} 
\author{Helena Pais} 
\email{hpais@uc.pt} 
\affiliation{CFisUC, Department of Physics, University of Coimbra, 3004-516 Coimbra, Portugal} 
\author{Constan\c{c}a Provid\^encia} 
\email{cp@uc.pt} 
\affiliation{CFisUC, Department of Physics, University of Coimbra, 3004-516 Coimbra, Portugal}

\begin{abstract}
In this work, we study the effect of strong magnetic fields on the equation of state (EoS) of warm, homogeneous, Neutron Star (NS) matter in beta equilibrium. NS matter is described within a relativistic mean field (RMF) approximation, including both models with non-linear meson terms or with density dependent nucleon-meson couplings. We first study the effect of magnetic fields and finite temperature on the EoS separately, finding that the effect of the latter to be significantly stronger than the one of the former. We then study the combined effect of magnetic fields and temperature on the internal composition. We show how both factors cause an increase in the proton fraction at low density and that, as long as the temperatures considered are not higher than 10 MeV, the effect of the magnetic field on the proton fraction is not small enough to be neglected.

\end{abstract}
\maketitle

\section{Introduction}

The EoS of NSs is a topic of extreme interest in the modern era of multi-messenger astronomy and represents a very complex challenge, due to the lack of data and the impossibility of reproducing their properties in ground-based experiments. Simulations that describe the formation or merging of these objects require the knowledge of the EoS over a wide range of temperatures, densities and proton fractions \cite{Oertel:2016bki}.

A category of NS, referred to as magnetars \cite{Duncan_1992,Thompson_1995,Usov_1992,Paczynski_1992}, show magnetic fields among the strongest observed in nature, with values at the surface up to $\sim 10^{16}$ G \cite{Olausen_2014,McGill}, and up to $\approx 10^{18}$ G in the core, according to both the scalar virial theorem \cite{Lai_1991,Shapiro_1983} and numerical solutions of the Maxwell-Einstein equations \cite{Chatterjee_2015,Gomes_2019,Cardall_2001,Sengo_2020,Bonazzola_1993,Bocquet_1995}.
The role of strong magnetic fields in the EoS has been studied in detail in the literature. Studies have shown that the effect of a strong magnetic field in the core EoS would only start to be seen for fields above $\sim 10^{18}$ G \cite{Chatterjee_2015,Broderick_2000}, however, there are direct magnetic fields on the macroscopic structure of the star that can not be neglected. In Gomes et al \cite{Gomes_2019}, the authors found that fields above $10^{17}$G in the core already give a non-negligible deformation of the star, and the full Maxwell-Einstein equations should be solved. The crust EoS, on the other hand, seems to feel more strongly the effect of an external magnetic field.
In \cite{Chamel_2012,Potekhin_2013,Chamel_2015,Stein_2016}, the authors study the effect of strong fields on the outer crust of NS. The effects on the inner crust have been studied using several approaches such as the calculation of the dynamical spinodal instability regions \cite{Rabhi_2009,Fang_2016,Fang_2017,Chen_2017,Fang_2017_2}, the coexisting phase or the compressible liquid drop approximations \cite{Pais_2021,Wang2022,Scurto_2023}, and self-consistent Thomas-Fermi calculations \cite{Mutafchieva_2019}. In these studies it was shown that, while the effect of the magnetic field on the EoS is indeed small, it can significantly affect the internal composition of NSs.

The EoS of NSs is often studied at zero temperature within various formalisms. However, a zero-temperature EoS is not valid in the case of proto-NS and other systems, such as core-collapse supernovae or binary NS mergers, because these objects evolve at temperatures that can be 50~MeV or higher. The effect of finite temperatures on the EoS has been studied using different approaches, such as RMF calculations \cite{Shen:1998gq,Prakash_1997,Menezes:2003xa,Hempel:2009mc,Pais_2015,Avancini_2017,Fortin:2017dsj,Kochankovski_2022,Guichon:2023iev}, Skyrme effective \cite{Lattimer:1991nc,Prakash_1997,Chamel_2009,Wei_2021,Raduta_2022} interactions, microscopic descriptions using many-body formalisms \cite{Liu:2022fgi,Benhar:2023mgk} and, more recently, calculations based on phenomenological nuclear Hamiltonians \cite{Tonetto2022}.

In this work we study the combined effect of strong magnetic fields and finite temperatures on the EoS of homogeneous stellar matter in beta equilibrium. The motivation for such an analysis comes from the fact that the magnetic field of NSs becomes weaker as the star ages, and thus the youngest and warmest stars are also expected to be the ones with the strongest magnetic fields. The effect of strong magnetic fields on warm stellar matter has already been studied in \cite{Rabhi2011}, where the authors use a RMF approximation like the one used in this work. 
However, in that work, the authors consider magnetic fields one order of magnitude stronger than the strongest one used in our study, as well as considering matter with a fixed charge fraction rather than in beta equilibrium. In Ref.~\cite{Fang_2017_2}, the authors studied the sub-saturation EoS using a thermodynamical spinodal calculation at finite temperatures and strong magnetic fields, and they found out that for temperatures above $\sim 100$keV, the effect of the magnetic field would be negligible.

In this study, stellar matter is described within a RMF approximation, and we compare the results obtained with four different RMF models: NL3 \cite{Lalazissis1997}, NL3$\omega\rho$ \cite{Horowitz2001,Pais2016}, EOS18 \cite{malik2024} and SPG(M4) \cite{scurto2024}. 
These models were selected to span different combinations of non-linear (NL) meson terms as well as density dependent couplings. 
The first three models include NL meson terms in the Lagrangian density, while the last one considers density dependent couplings for the nucleon-meson interaction. The NL3 and the NL3$\omega\rho$ models share the same isoscalar properties. The latter was built because NL3 has a very large slope of the symmetry energy at saturation, $L=118$ MeV. In the NL3$\omega\rho$ model, an interaction term between the $\omega$ and the $\rho$ meson is added to model the density dependence of the symmetry energy and to lower its slope at saturation, $L=55$ MeV. The EOS18 was part of a dataset used in \cite{Malik_2023} and was then selected in \cite{malik2024}, as part of a few EoS that span a large range of the slope of the symmetry energy. This particular model came out to be in good agreement with both experimental observations and theoretical predictions. This EoS includes both 
 the $\omega\rho$ interaction term, present also in NL3$\omega\rho$, and a quartic self-interaction term for the $\omega$ meson. Finally, SPG(M4) was part of a dataset used in \cite{scurto2024}, where the authors perform a Bayesian analysis with unified RMF density dependent EoSs, and was selected, together with four other models, chosen in order to ensure as much variety as possible and which included SPG(M4), to be uploaded on the public repository CompOSE.

The paper is structured as follows: in Section \ref{RMFf}, we present the formalism in detail, with a particular focus on the role of the magnetic field and temperature. In Section \ref{results}, we present and analyse the results of the study. Finally, in Section \ref{conclusions}, we draw some conclusions. 

\section{Relativistic Mean Field formalism \label{RMFf}}

Here we present the theoretical framework of our study. We start by summarizing the RMF description of the EoS in the case of cold and non-magnetized matter. We then show how the calculation is modified by the presence of finite temperature and finite magnetic field separately. Finally, we show the case in which both temperature and magnetic field are different from zero. 
In all the cases studied, we consider npe$\mu$ matter described within a RMF approximation, where the interaction between the nucleons is mediated by three types of mesons: the isoscalar-scalar meson $\sigma$, the isoscalar-vector meson $\omega$ and the isovector-vector meson $\rho$. Charge neutrality and the weak equilibrium conditions are imposed in all the considered cases:
\begin{align}
 &\mu_n-\mu_p=\mu_e \; ; \; \mu_e=\mu_\mu \, , \\
 &\rho_p=\rho_e+\rho_\mu \, .
\end{align}

\subsection{Cold Non Magnetized Matter}

For $B=0$ and $T=0$, the Lagrangian density of our system is given by

\begin{equation}    \mathcal{L}=\sum_{i=p,n}\mathcal{L}_i+\mathcal{L}_l+\mathcal{L}_\sigma+\mathcal{L}_\omega+\mathcal{L}_\rho+\mathcal{L}_{int}\, ,
\end{equation}
where $\mathcal{L}_l$ is the standard lepton Lagrangian density  given by
\begin{equation}
    \mathcal{L}_l=\sum_{l=e,m}\bar{\psi}_l\big[\gamma_\mu i\partial^\mu-m_l\big]\psi_l \, ,
\end{equation}
the nucleon Lagrangian density is given by 
\begin{equation}
    \mathcal{L}_i=\bar{\psi}_i\big[\gamma_\mu iD^\mu-M_*\big]\psi_i \, ,
\end{equation}
with
\begin{equation}
    M_*=M-g_\sigma\phi \, ,
\end{equation}
and
\begin{equation}
    iD^\mu=i\partial^\mu-g_\omega V^\mu-\frac{g_\rho}{2}\mathbf{\tau}\cdot\mathbf{b}^\mu \, ,
\end{equation}
where $\tau$ are the Pauli matrices, so that the third component is $\tau_3=\pm 1$ for protons and neutrons respectively.
The mesonic components of the Lagrangian density  are expressed as
\begin{equation}
    \mathcal{L}_\sigma=\frac{1}{2}\bigg(\partial_\mu\phi\partial^\mu\phi-m_\sigma^2\phi^2 \bigg) \, ,
\end{equation}
\begin{equation}
    \mathcal{L}_\omega=-\frac{1}{4}\Omega_{\mu\nu}\Omega_{\mu\nu}+\frac{1}{2}m_\omega^2V_\mu V^\mu  \, ,
\end{equation}
\begin{equation}
    \mathcal{L}_\rho=-\frac{1}{4}\mathbf{B}_{\mu\nu}\cdot\mathbf{B}^{\mu\nu}+\frac{1}{2}m_\rho^2\mathbf{b}_\mu\cdot \mathbf{b}^\mu \, ,
\end{equation}
with the tensors  
defined as
\begin{eqnarray}
\Omega_{\mu\nu}&=&\partial_\mu V_\nu - \partial_\nu V_\mu \, , \\
\mathbf{B}_{\mu\nu}&=&\partial_\mu \mathbf{b}_\nu - \partial_\nu \mathbf{b}_\mu - g_\rho \left(\mathbf{b}_\mu \times \mathbf{b}_\nu \right)\, .
\end{eqnarray}

The NL models, i.e. NL3, NL3$\omega\rho$, and EOS18, include the following higher order meson-meson interaction terms, given by
\begin{align}
    \mathcal{L}_{int}=\frac{\xi}{4!} &g_\omega^4(V_\mu V^\mu)^2-\frac{1}{6}\kappa\phi^3\\
    &-\frac{1}{24}\lambda\phi^4+\Lambda_{\omega\rho}g_\omega^2g_\rho^2V_\mu V^\mu\mathbf{b}_\mu\cdot \mathbf{b}^\mu \, , \notag
\end{align}
while SPG(M4) considers density dependent couplings, where each of the three nucleon-meson couplings is given by the GDFM density functional \cite{Gogelein:2007qa} 
\begin{equation}
    g_i=a_i+(b_i+d_ix^3)e^{-c_i x}\, , \label{eq:couplings}
\end{equation}
where $x=\rho/\rho_{sat}$, with $\rho_{sat}$ being the saturation density of the model.

The particle number densities are given by 
\begin{equation}
    \rho_i=\frac{{k_i^{F}}^3}{3\pi^2} \, ,
\end{equation}
with $i=p,n,e,\mu$ and the Fermi momentum is 
\begin{equation}
    k_i^F=\sqrt{\eta_i^2-m^2}\, ,
    \label{Eq.fermi_mom}
\end{equation}
where $\eta_i$ is the effective chemical potential for nucleons and the electron chemical potential for leptons, and $m=M_*$ for nucleons and $m=m_l$ for leptons.

The chemical potentials are defined as 
\begin{align}
    \mu_p=&\eta_p+g_\omega V^0+\frac{1}{2}g_\rho b^0+\Sigma_R \, , \\
    \mu_n=&\eta_n+g_\omega V^0-\frac{1}{2}g_\rho b^0+\Sigma_R \, .
\end{align}
The rearrangement term $\Sigma_R$ only appears in the SPG(M4) model and is given by 
\begin{align}
    \Sigma_R(\rho)=\frac{\partial g_\omega}{\partial\rho}\omega_0\rho+\frac{1}{2}\frac{\partial g_\rho}{\partial \rho}b_{0,3} \rho_3-\frac{\partial g_\sigma}{\partial \rho}\phi_0\rho_s \, .
\end{align}

The energy density and pressure are defined respectively by 
\begin{equation}    \mathcal{E}=\mathcal{E}_{kin}^p+\mathcal{E}_{kin}^n+\mathcal{E}_{kin}^l+\mathcal{E}_F \, ,
\end{equation}
\begin{equation}   
P=P_{kin}^p+P_{kin}^n+P_{kin}^l+P_F \, ,
\end{equation}
where kinetic terms, $\mathcal{E}_{kin}^i$ and $P_{kin}^i,\, i=p,n,l$  are written as 
\begin{equation}
    \mathcal{E}_{kin}^i= \frac{1}{\pi^2}\int_0^{k_i^F} dk\,k^2 \sqrt{k^2+m^{2}},
\end{equation}
\begin{equation}
    P_{kin}^i=\frac{1}{3\pi^2}\int_0^{k_{i}^F} dk\,\frac{k^4}{ \sqrt{k^2+m^{2}}},
\end{equation}

where again $m=M_*$ for nucleons and $m=m_l$ for leptons.
The field contributions are given by 

\begin{align}
    \mathcal{E}_{F}=\frac{m_\omega^2}{2}\omega_0^2&+\frac{\xi g_\omega^4}{8}\omega_0^4+\frac{m_\rho^2}{2}b_{3,0}^2+\frac{m_\sigma^2}{2}\phi_0^2 \nonumber \\
+&\frac{\kappa}{6}\phi_0^3+\frac{\lambda}{24}\phi_0^4+3\Lambda_{\omega\rho} g_\rho^2g_\omega^2\omega_0^2b_{3,0}^2
\end{align}
\begin{align}
\mathcal{P}_{F}=&\frac{m_\omega^2}{2}\omega_0^2+\frac{\xi g_\omega^4}{24}\omega_0^4+\frac{m_\rho^2}{2}b_{3,0}^2-\frac{m_s^2}{2}\phi_0^2 \nonumber \\
&-\frac{\kappa}{6}\phi_0^3-\frac{\lambda}{24}\phi_0^4+\Lambda_{\omega\rho} g_\rho^2g_\omega^2\omega_0^2b_{3,0}^2+\rho\Sigma_R.
\end{align}

\subsection{Cold Magnetized Matter}

We now introduce the effect of the magnetic field, following previous studies \cite{Rabhi_2015,Pais_2022,Wang2022,Scurto_2023}. We consider an electromagnetic field 
of the type $A^\mu=(0,0,Bx,0)$, so that the resulting field is oriented along the $z$ axis. We do not include the anomalous magnetic moment of particles, since it was shown in previous studies \cite{Wang2022} that its contribution is smaller than the already small correction given by the interaction between the magnetic field and charged particles. 

In the following, we will refer to the quantity $B^*$, defined as $B^* = B/B^c_e$, with $B^c_e= 4.414 \times 10^{13}$ G being the critical field at which the electron cyclotron energy is equal to the electron mass.

In the presence of a magnetic field, the nucleon and lepton terms of the Lagrangian density become respectively
\begin{eqnarray}
    \mathcal{L}_i&=&\bar{\psi}_i\big[\gamma_\mu iD^\mu-M_*\big]\psi_i \, , i=p,n \, , \\
    \mathcal{L}_l&=&\sum_{l=e,\mu}\bar{\psi}_l\big[\gamma_\mu\big(i\partial^\mu+eA^\mu\big)-m_l\big]\psi_l \, , 
\end{eqnarray}

with
\begin{equation}
    iD^\mu=i\partial^\mu-g_\omega V^\mu-\frac{g_\rho}{2}\mathbf{\tau}\cdot\mathbf{b}^\mu-\frac{1+\tau_3}{2}eA^\mu \, ,
\end{equation}
where $e=\sqrt{4\pi/137}$ is the electron charge. 
Moreover, an extra term for the electromagnetic field appears in the Lagrangian density: 
\begin{equation}
    \mathcal{L}_A=-\frac{1}{4}F_{\mu\nu} F^{\mu\nu} \, ,
\end{equation}
with $F_{\mu\nu}=\partial_\mu A_\nu-\partial_\nu A_\mu$.\\
Since we neglect the anomalous magnetic moment of particles, all quantities for neutrons remain unchanged with respect to the $B=0$ case. The proton and lepton densities become  
\begin{equation}
    \rho_i=\frac{|q|B}{2\pi^2}\sum_{\nu=0}^{\nu_{\rm max}^i} g_\nu k^F_{i,\nu} \, ,
\end{equation}

with $i=p,l$, where $\nu=n+\frac{1}{2}-\frac{1}{2}\frac{q}{|q|}s=0,1,\cdots,\nu_{\rm max}$ enumerates the Landau levels (LLs) for fermions with electric charge $q$. The spin degeneracy factor of the LLs, $g_\nu$, is equal to $g_\nu=1$ for $\nu=0$ and $g_\nu=2$ for $\nu>0$, $\nu_{\rm max}$ is the largest LL 
occupied by fully degenerate charged fermions, defined as

\begin{align}
    \nu_{\rm max}^i=\frac{\eta_i^{2}-m^{2}}{2 |q| B} \, .  
\end{align}
and the Fermi momentum for charged particles is given by
\begin{align}
    k^{F}_{i,\nu}=&\sqrt{\eta_i^{2}-m^{2}- 2\nu |q| B} \,.
    \label{Eq.fermi_mag}
\end{align}
where $\eta_i$ and $m$ are defined as in Eq.\ref{Eq.fermi_mom}~.

The proton and lepton kinetic terms in the equation for the energy density become

\begin{align}
    \mathcal{E}_{kin}^i=\frac{|q|B}{4\pi^2}\sum_{\nu=0}^{\nu_{\rm max}}g_s\bigg[k_{F,\nu}^i\eta_i&+\bigg(m^{2}+2\nu |q|B \bigg)\cdot \notag\\&\ln\bigg|\frac{k_{F,\nu}^i+\eta_i}{\sqrt{m^{2}+2\nu |q|B}}\bigg|\bigg] \, , 
\end{align}

where $\eta_i$ and $m$ are defined as in Eq.\ref{Eq.fermi_mom}~ and the Fermi momentum is defined as in Eq.\ref{Eq.fermi_mag}~.

In this case, we calculate the pressure from the thermodynamic relation

\begin{equation}
   P=\mu_p\rho_p+\mu_n\rho_n+\mu_e(\rho_e+\rho_\mu)-\mathcal{E}.
\end{equation}

\subsection{Warm Non Magnetized Matter}

We proceed by showing the effect of finite temperature on the RMF formalism in the absence of an external magnetic field. In this case, the Lagrangian density of the system is the same as the zero-temperature case,
the only difference being the occupation number distribution functions for the different particles, given by

\begin{equation}
    f^i_{k,\pm}=\frac{1}{1+\exp(\frac{\epsilon_k^i\pm \eta_i}{T})} \, , i=p,n,e,\mu \, .
\end{equation}

The minus sign is for particles and the plus is for anti-particles and $\eta_i$ is the effective chemical potential in the case of nucleons and the chemical potential in the case of leptons. In the previous equation we have
\begin{equation}
    \epsilon_k^i=\sqrt{k^2+m^2} \, ,
\end{equation}
with $m=M_*$ for nucleons and $m=m_l$ for leptons. \\
The number density is thus given by
\begin{align}
    \rho_i&=2\int \frac{d^3k}{(2\pi)^3}(f^i_{k,-}-f^i_{k,+})  \notag \\
    &=\frac{1}{\pi^2}\int_0^\infty k^2(f^i_{k,-}-f^i_{k,+}) dk \, ,
\end{align}
where the second equality holds because the integrand only depends on the modulus of the momentum. In our work, the Fermi integrals are evaluated using the method developed in \cite{Aparicio_1998}.

The kinetic contributions to the energy density and pressure are  given by  
\begin{align}
    \mathcal{E}_{kin}^i&=\frac{1}{\pi^2}\int_0^\infty \epsilon_k^i k^2(f^i_{k,-}+f^i_{k,+})dk,
\end{align}
and 
\begin{align}
    P_{kin}^i&=\frac{1}{3\pi^2}\int_0^\infty  \frac{k^4}{\epsilon_k^i}(f^i_{k,-}+f^i_{k,+})dk \, .
\end{align}

\subsection{Warm Magnetized Matter}

We end the description of our theoretical framework by showing the case in which both finite temperature and finite magnetic field are taken into account. The same formalism was already used in previous studies \cite{Rabhi2011}.

As in the case with finite $B$ and $T=0$, we neglect the anomalous magnetic moment of the particles, so that all the quantities concerning the neutrons remain the same as in warm non-magnetized case.

In this case, the occupation number distribution functions for protons and leptons is
\begin{equation}
    f^i_{k,\nu,\pm}=\frac{1}{1+exp(\frac{\epsilon_{k,\nu}^i\pm \eta_i}{T})} \, ,
\end{equation}
where 
\begin{equation}
    \epsilon_{k,\nu}^i=\sqrt{k_z^2+m^2+2\nu |q|B} \, ,
\end{equation}
with $m=M_*$ for protons and $m=m_l$ for leptons. We notice that, in the presence of the magnetic field, $\epsilon_{k,\nu}^i$ directly depends on the magnetic field and on the LL $\nu$, and moreover it depends on the component of the momentum parallel to the field, rather than on its modulus.

The number density for the charged particles is given by 
\begin{equation}
    \rho^i=\sum_\nu \rho^i_\nu=\sum_\nu \frac{g_\nu |q|B}{2\pi^2}\int _0^\infty (f_{k,\nu,-}^i-f_{k,\nu,+}^i)dk_z \, ,
\end{equation}
where $g_\nu$ is the spin degeneracy of the LLs. Here, unlike in the cold magnetized case, the summation is done over all LL. In the numerical calculation, the sum is truncated when the occupation of an LL becomes several orders of magnitudes smaller than the one of the first level. This dynamical estimation of the cut-off allows us to be sure not to cut the summation too early, especially in the high $T$ and low $B$ cases, in which the number of levels that need to be taken into account can easily reach the order of $\sim 10^3$.

The kinetic contributions to the energy density and pressure for charged particles are given respectively by 
\begin{equation}
    \mathcal{E}^i=\sum_\nu \frac{g_\nu |q|B}{2\pi^2}\int _0^\infty \epsilon_{k,\nu}^i (f_{k,\nu,-}^i+f_{k,\nu,+}^i)dk_z \, ,
\end{equation}
and 
\begin{align}
    P^i=\sum_\nu \frac{g_\nu |q|B}{2\pi^2}\int _0^\infty \frac{k_z^2}{\epsilon_{k,\nu}^i}(f_{k,\nu,-}^i+f_{k,\nu,+}^i)dk_z \, . 
\end{align}

We finally introduce a notation that will be used in the next section. We define
\begin{equation}
    \Delta_B X \equiv \frac{X_{B\neq 0}-X_{B= 0}}{X_{B= 0}}\Bigg\vert_T \, ,
    \label{B_var}
\end{equation}
for the difference between a quantity $X$ evaluated at finite B and the same quantity evaluated at $B=0$, normalized by the quantity evaluated at $B=0$, fixing the temperature, and 
\begin{equation}
    \Delta_T X \equiv \frac{X_{T\neq 0}-X_{T= 0}}{X_{T= 0}}\Bigg\vert_B \, ,
    \label{T_var}
\end{equation}
for the difference varying the temperature, while keeping the magnetic field fixed.

\begin{table*}
    \begin{tabular}{ccccccc}
    \hline
    \hline
          & $B/A$ (MeV) & $\rho_0$ (fm$^{-3}$) & $M^*/M$ & $K$ (MeV) & $\mathcal{E}_{sym}$ (MeV) & $L$ (MeV)  \\
    \hline
        NL3 & -16.24 & 0.148 & 0.60 & 270 & 37.34 & 118 \\
        NL3$\omega\rho$ & -16.24 & 0.148 & 0.60 & 270 & 31.66 & 55\\
        EoS18 & -16.31 & 0.155 & 0.67 & 221 & 32 & 49\\
        SPG(M4) & -15.8 & 0.162 & 0.62 & 246 & 30.9 & 40.1\\
    \hline
    \hline
    \end{tabular}
    \caption{Symmetric nuclear matter properties at saturation density for the four models used in our analysis. From left to right:  binding energy per baryon, saturation density, normalized nucleon effective mass, incompressibility, symmetry energy and slope of the symmetry energy.}
    \label{Tab1}
\end{table*} 

\section{Results \label{results}}

We now proceed to illustrate the results of our study.

In Tab.\ref{Tab1} we show the properties of symmetric nuclear matter at saturation for the four selected models. It can be seen that the isovector properties of NL3 are significantly larger as compared to the other three models, which present similar values for both the symmetry energy and its slope at saturation density.

\begin{figure}
  \centering
  \includegraphics[width=\linewidth,angle=0]{ 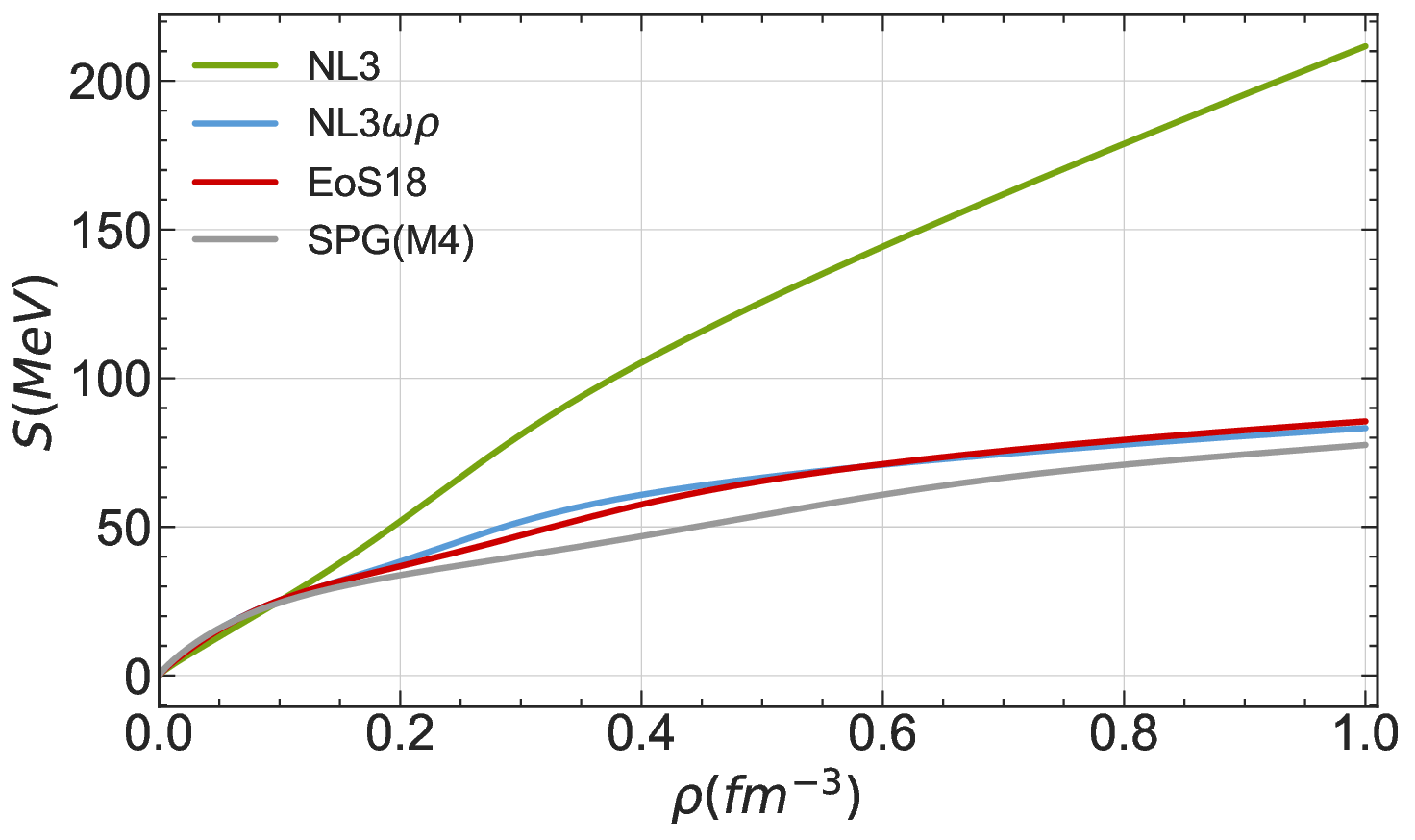}  
\caption{Symmetry energy as a function of the baryon density for the four considered models.}
    \label{Fig_E_sym}
\end{figure}

This can also be observed in Fig.\ref{Fig_E_sym}, where we plot the symmetry energy as a function of the baryonic density. This difference in the behaviour of the symmetry energy explains the different behaviour of the NL3 model in the following results. Note, however, that below $\sim 0.1$fm$^{-3}$ the symmetry energy of NL3 is the lowest, and this property will also distinguish this model from the others.

\begin{figure}
  \centering
  \includegraphics[width=\linewidth,angle=0]{ 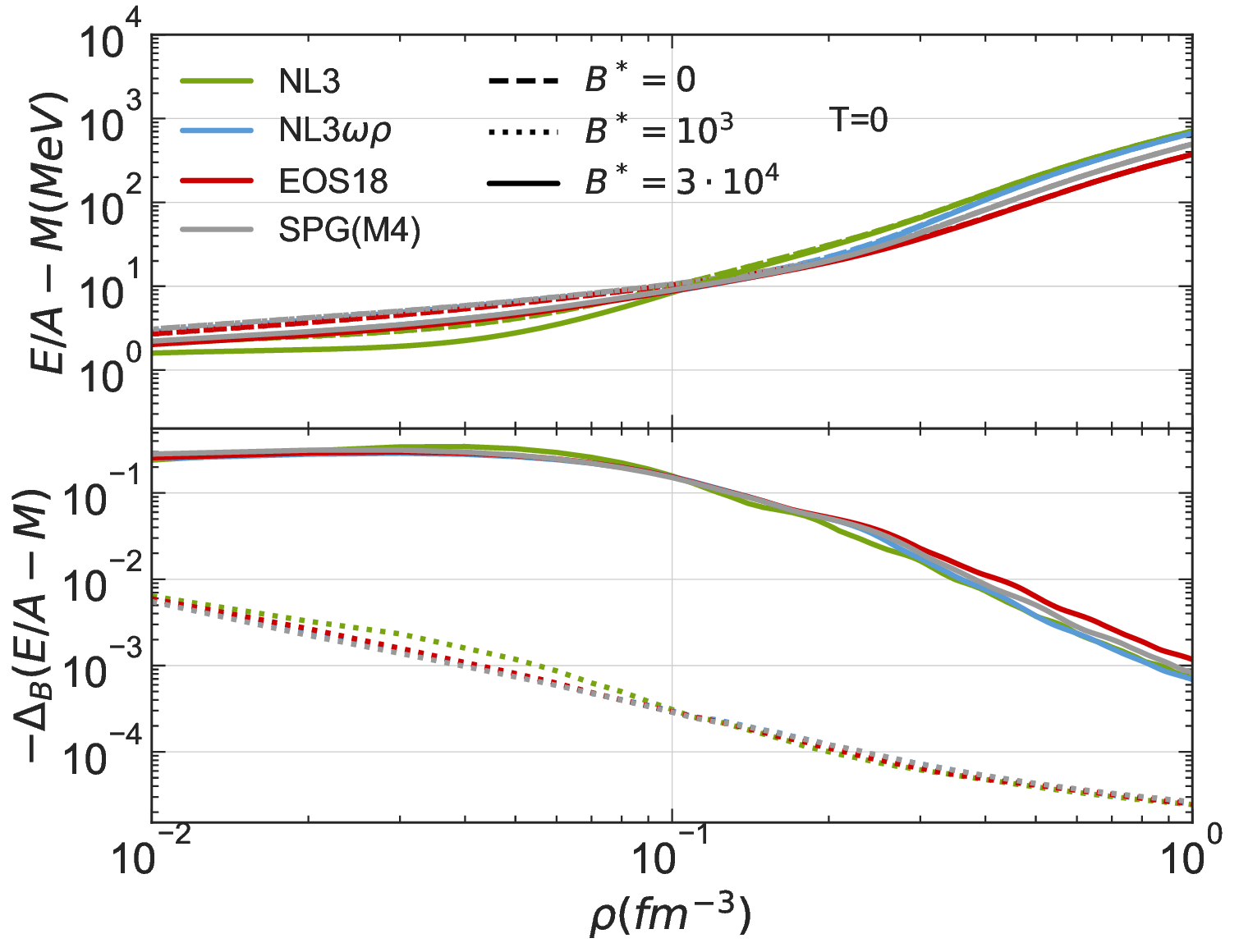}  
\caption{Energy per baryon (top) and its relative variation with respect to the zero-$B$ solution (as defined in Eq.~\ref{B_var}) (bottom) as a function of the density. The results for $B^*=0$ (dashed), $B^*=10^3$ (dotted), and $B^*=3\cdot 10^4$ (solid) and for the four models are shown.}
    \label{Fig_E_B}
\end{figure}

We start our analysis by studying the effect of strong magnetic fields at zero temperature. In Fig.~\ref{Fig_E_B}, we show in the upper panel the energy per baryon as a function of the baryonic density for different values of the effective magnetic field $B^*$ and in the lower panel we plot the symmetric of the relative variation of the energy per baryon with the magnetic field with respect to the zero-field solution, as defined in Eq.~\ref{B_var}. We see that the magnetic field has the effect of reducing the energy per baryon. However, the effect of the magnetic field is only a small correction, since the variation is less than 1\% for the lowest value of the field considered ($B^*=10^3$, i.e., $B=4.4\times 10^{16}$G). For the highest field ($B^*=3\times 10^4$, i.e. $B=1.32\times 10^{18}$G)  considered, it becomes less than 10\%  above saturation density, but below this density the effect can be as large as 40\%-50\%. It is also seen that the effect of the field decreases rapidly with density and is only relevant at lower densities.

\begin{figure}
  \centering
  \includegraphics[width=\linewidth,angle=0]{ 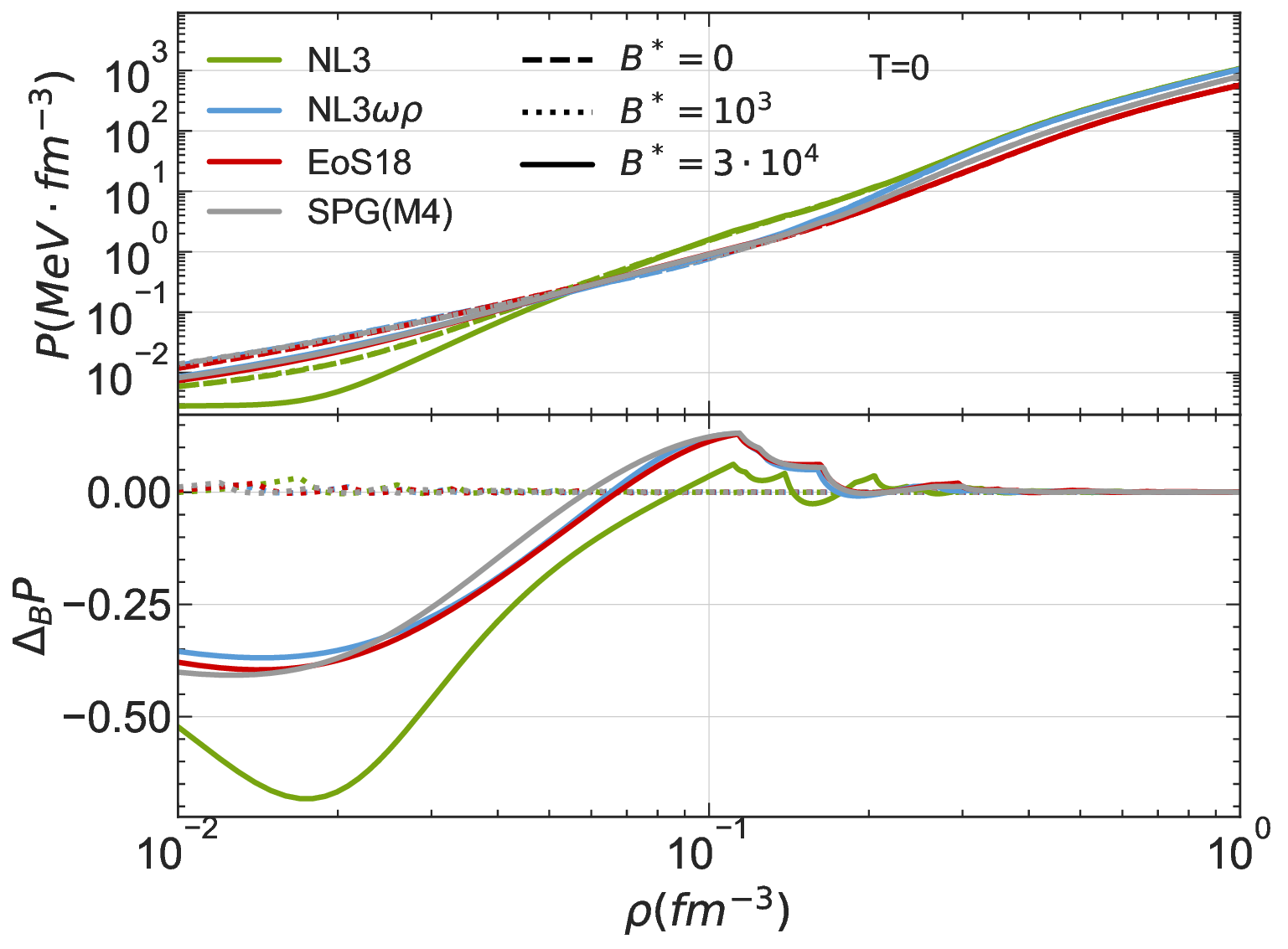}  
\caption{Pressure (top) and its relative variation with $B$ (as defined in Eq.~\ref{B_var}) (bottom) as a function of the density. The results for $B^*=0$ (dashed), $B^*=10^3$ (dotted) and $B^*=3\cdot 10^4$ (solid) and for the four models are shown.}
    \label{Fig_P_B}
\end{figure}

\begin{figure}
  \centering
  \includegraphics[width=\linewidth,angle=0]{ 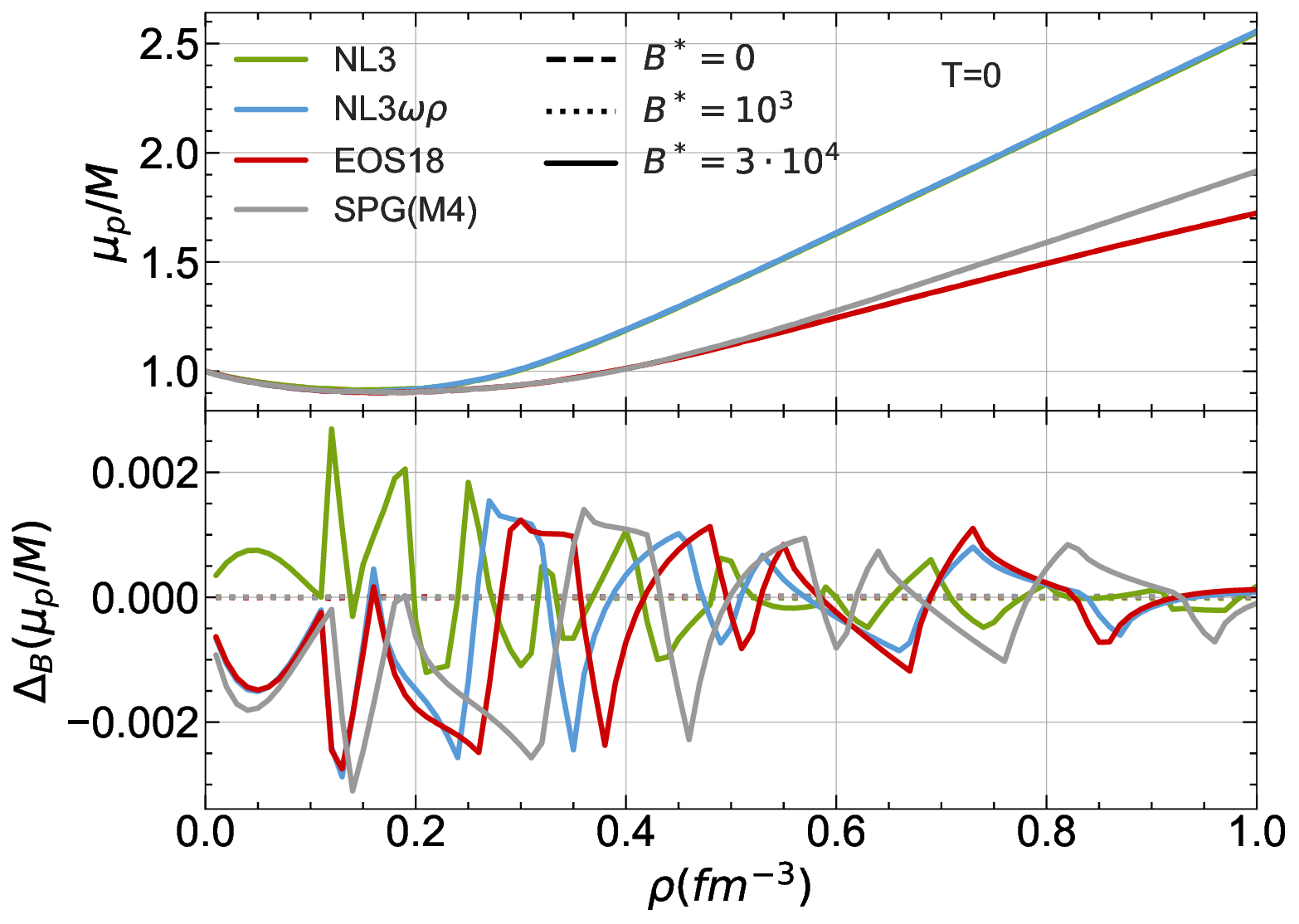}  
\caption{Normalized proton chemical potential (top) and its relative variation with $B$ (as defined in Eq.~\ref{B_var}) (bottom) as a function of the density. The results for $B^*=0$ (dashed), $B^*=10^3$ (dotted) and $B^*=3\cdot 10^4$ (solid), and for the four models are shown.}
\label{Fig_Mu_B}
\end{figure}

In Fig.~\ref{Fig_P_B} we show the analogue of the previous plot for the pressure. In this case, the variation due to the field oscillates between positive and negative values. This behaviour is due to the presence of the LLs and, in the case of the highest field, it is possible to see in the plot the points where a new level opens. However, as in the case of the energy per baryon, also in the case of the pressure, the correction due to the presence of the field is small, and decreases rapidly with density, so it is only relevant at low densities and only for very high values of the field. The effect at low densities appears to be largest for NL3 because in this range of densities, the model has the smallest symmetry energy and $\beta$-equilibrium matter has a very small fraction of protons.  

Similar conclusions apply to the chemical potential of the proton, shown in Fig.~\ref{Fig_Mu_B}. Again, the variation oscillates between positive and negative values due to the Landau quantization. However, the effect of the field is even smaller than in the case of pressure, being of the order of $\sim 10^{-3}$ for the highest value of the field considered, and significantly smaller for the lowest, so that it always appears to be zero in the plot.

\begin{figure}
  \centering
  \includegraphics[width=\linewidth,angle=0]{ 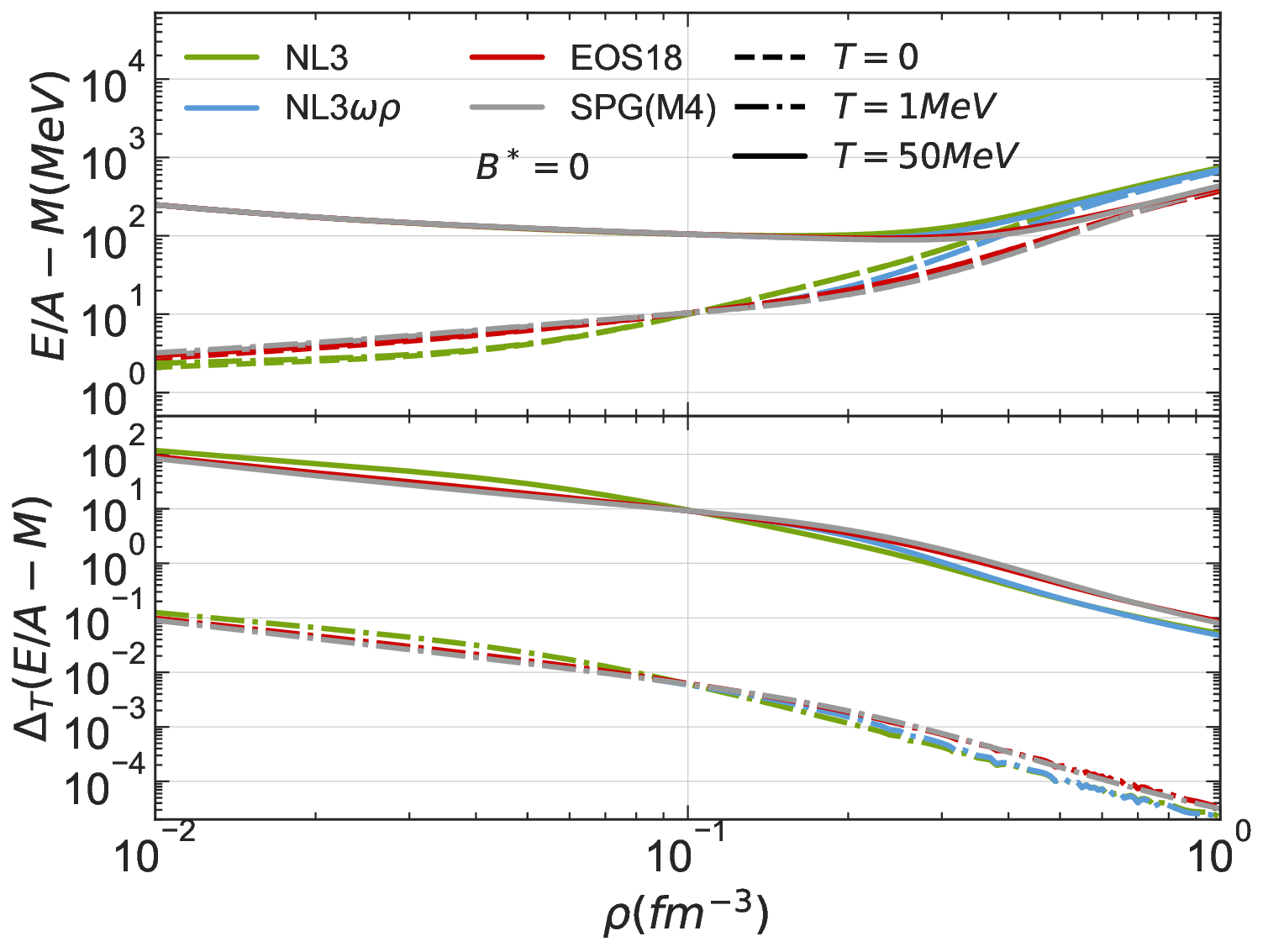}  
\caption{Energy per baryon (top) and its variation with $T$ (as defined in Eq.~\ref{T_var}) (bottom) as a function of the density. The results for $T=0$ (dashed), $T=1$ MeV (dotted) and $T=50 $~MeV (solid), and for the four models are shown.}
\label{Fig_E_T}
\end{figure}

\begin{figure}
  \centering
  \includegraphics[width=\linewidth,angle=0]{ 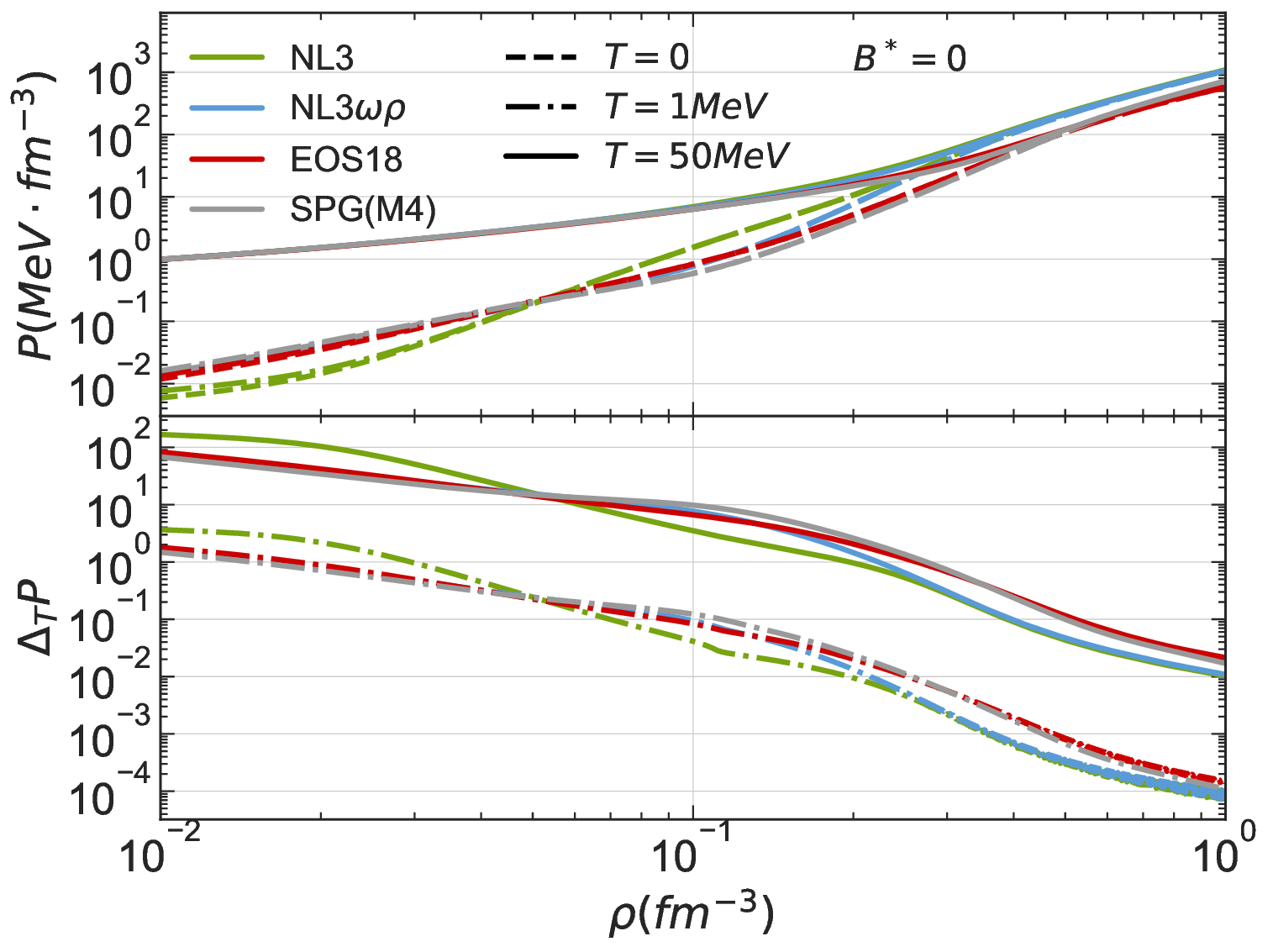} 
  \caption{Pressure (top) and its variation with $T$ (as defined in Eq.~\ref{T_var}) (bottom) as a function of the density. The results for $T=0$ (dashed), $T=1 $~MeV (dotted) and $T=50 $~MeV (solid), and for the four models are shown.}
    \label{Fig_P_T}
\end{figure}

We now compare the effect of magnetic fields on the EoS with the effect of finite temperatures. In Figs.~\ref{Fig_E_T} and \ref{Fig_P_T}, we show the same quantities as in Figs.~\ref{Fig_E_B} and \ref{Fig_P_B}, but this time we consider non-magnetized matter, and examine its variation with temperature. 

Finite temperatures have the effect of increasing both the energy per baryon and the pressure, with a stronger effect at lower densities and for higher temperatures, but with a significant variation already at low temperatures ($T=1$ MeV). This result is in good agreement with previous studies 
\cite{Tonetto2022,Pais_2015}. Moreover, we can see that the effect of the temperature is significantly larger than that of the magnetic field, both for the energy per baryon and for the pressure, already for low temperatures ($T=1$ MeV).

\begin{figure}
  \centering
  \includegraphics[width=\linewidth,angle=0]{ 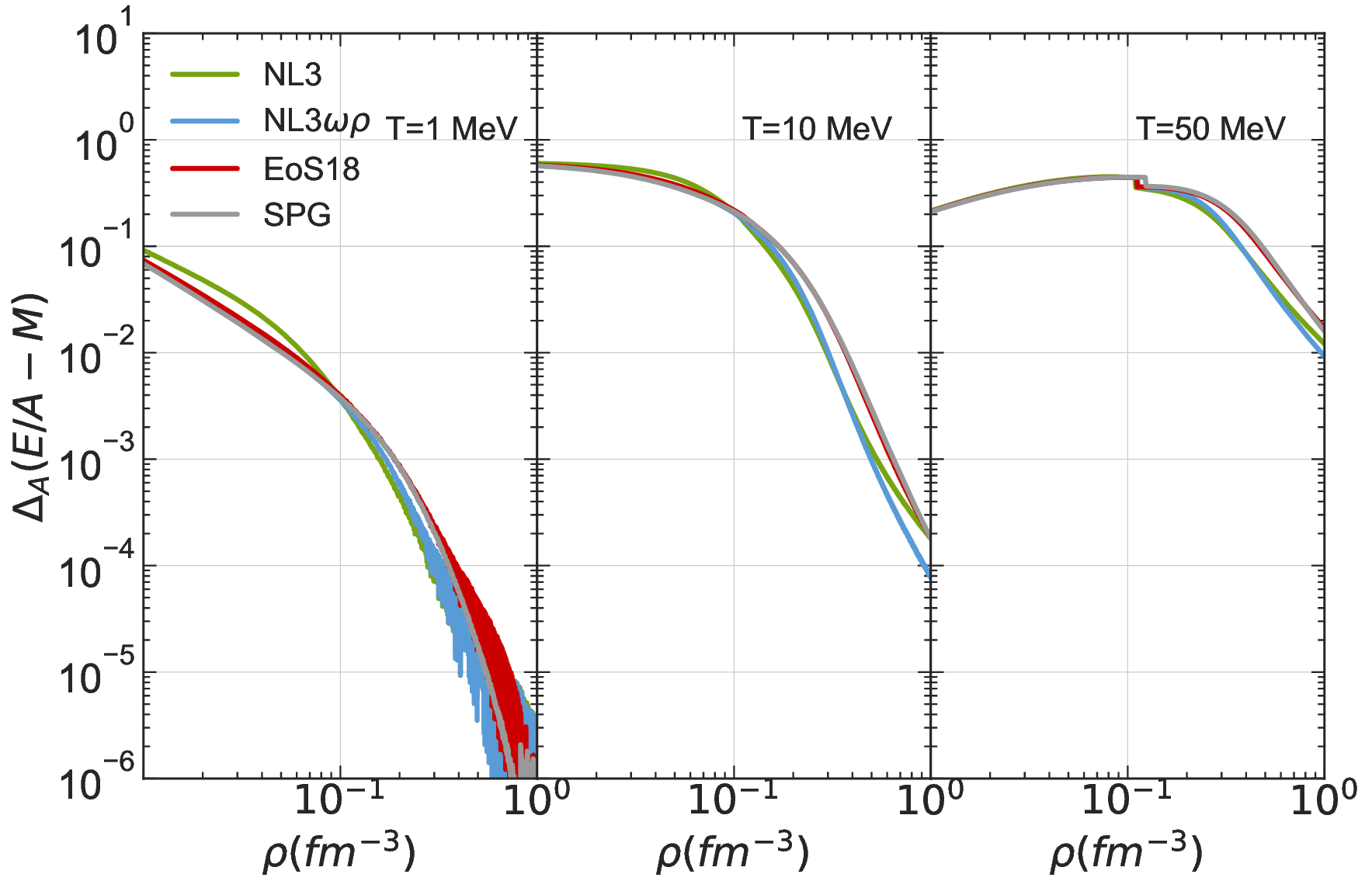} 
  \caption{Difference between the energy per baryon calculated with the RMF formulation and the one calculated with the approximation presented in \cite{Raithel_2019}, as defined in Eq.\ref{Eq.raithel}, for $B^*=0$. The results for $T=1 $~MeV (left panel), $T=10$~MeV (central panel) and for $T=50$~MeV (right panel) and for the four models are shown.}
    \label{Fig_Approx}
\end{figure}

Before discussing the effect of temperature and magnetic field on the composition of NSs, we compare our results at finite $T$ and $B^*=0$ with the approximation introduced in \cite{Raithel_2019}, where the authors propose a simple approximation to estimate the temperature contribution to the cold EoS. In Fig.~\ref{Fig_Approx}, we show the difference between the energy per baryon calculated with our approach and the one obtained using Eq. (26) of \cite{Raithel_2019} as a function of density, and for three values of the temperature. The difference is calculated as
\begin{equation}
    \Delta_A X= \frac{X_{RMF}-X_{App}}{X_{RMF}} \, ,
    \label{Eq.raithel}
\end{equation}
where $X_{RMF}$ is the quantity calculated with the RMF approximation and $X_{App}$ is the one obtained using the approximation. The approximation appears to be more precise at low temperatures, with a maximum error smaller than 10\% for $T=1$ MeV and $\rho\sim 0.01$~fm$^{-3}$. Still for $T=1$~MeV but at higher densities, where we already have shown that the temperature contribution is smaller, the relative variation may be as small as  10$^{-5}$. For $T\gtrsim 10$~MeV the relative difference becomes larger and takes values well above 10\% below saturation density. For a given temperature the approximation works always better at large densities.

\begin{figure}
  \centering
  \includegraphics[width=\linewidth,angle=0]{ 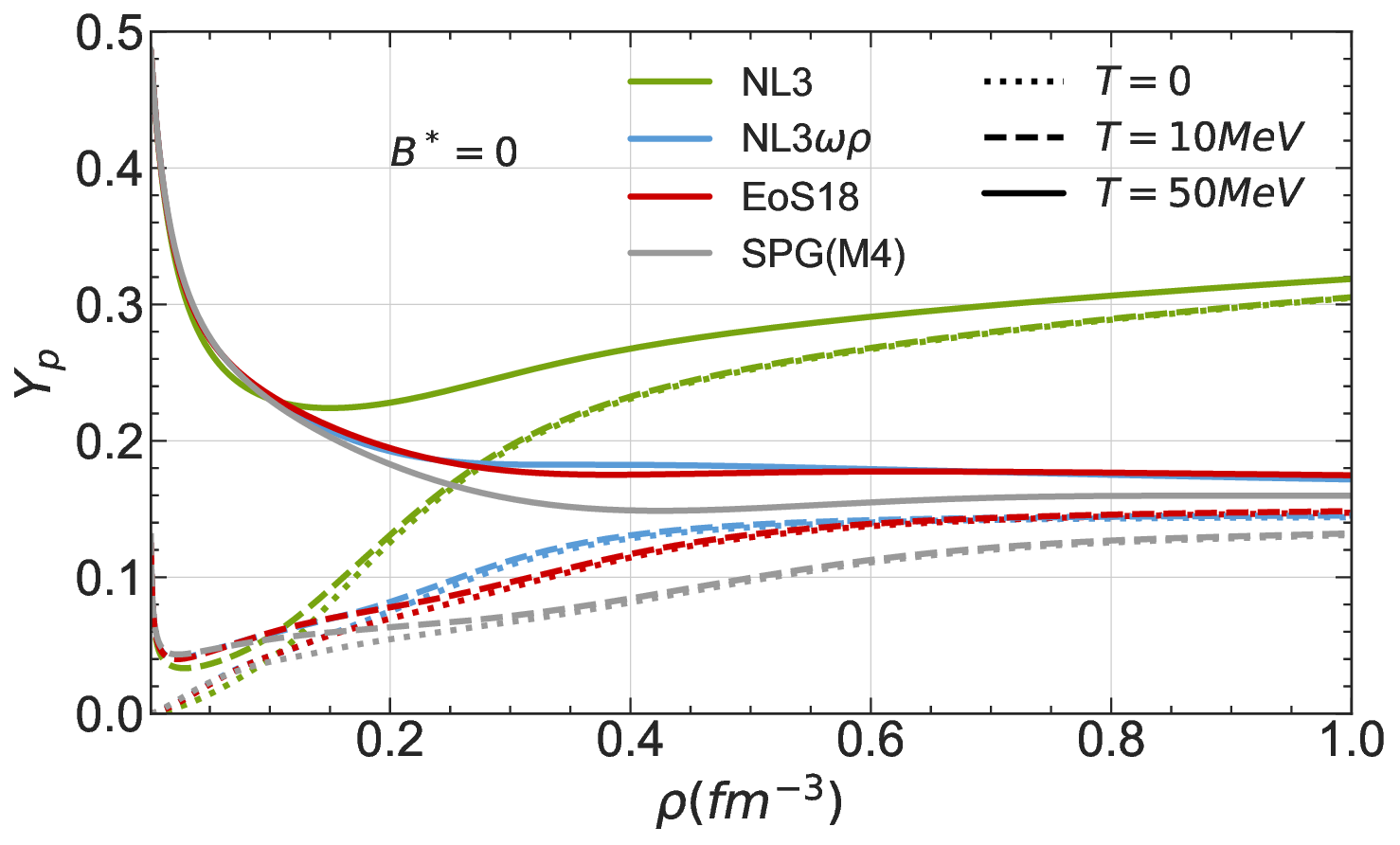} 
  \caption{Proton fraction as a function of density for $T=0$ (dotted), $T=10$~MeV (dashed) and $T=50$~MeV (solid) and for the four models are shown.}
    \label{Fig_Yp_T}
\end{figure}

We now focus on the effect of temperature and magnetic field on the composition of beta-equilibrated matter. The effect of the magnetic field has already been studied in previous studies \cite{Wang2022,Scurto_2023}, where it was shown that the main effect of the field is to increase the proton fraction at low densities and to introduce kinks in the proton fraction due to the Landau quantization. In Fig.~\ref{Fig_Yp_T}, we show the effect of finite temperatures on the proton fraction. Like the magnetic field, the temperature also has the effect of increasing the proton fraction at low densities, reaching almost symmetric matter at very high temperatures ($T=50$ MeV). This is in good agreement with what was shown in \cite{Tonetto2022}. We also note that for this value of temperature, the proton fraction remains higher than in the $T=0$ case even at very high densities. This effect can be as large as 15-30\% at 4$\rho_0$.

\begin{figure}
  \centering
  \includegraphics[width=\linewidth,angle=0]{ 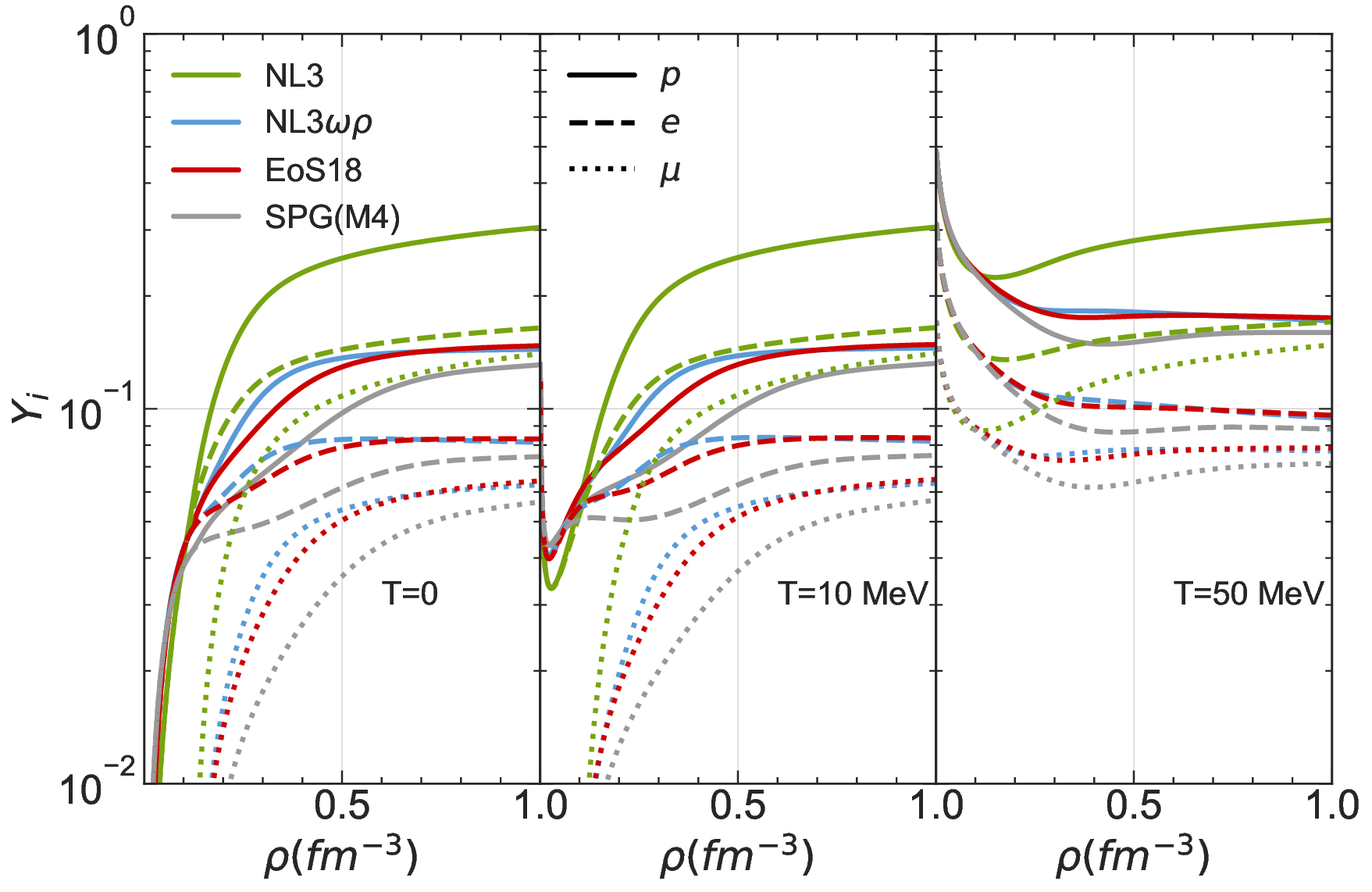} 
  \caption{Charged particle ($p,\, e,\, \mu$) fractions as a function of density for the four models and for $B^*=0$. The results for $T=0 $~MeV (left panel), $T=10$~MeV (central panel) and for $T=50$~MeV (right panel) are shown.}
    \label{Fig_Yi_T}
\end{figure}

The increase in the proton fraction caused by high temperatures also affects the lepton fraction and the density at which muons appear.  This is shown in Fig.~\ref{Fig_Yi_T}, where we plot the fraction of charged particles as a function of density for three values of the temperature. In agreement with \cite{Kochankovski_2022}, we show that for sufficiently high temperatures, muons begin to appear already at very low densities. In particular, at T=50 MeV, the muon fraction at low densities, smaller than 0.1 fm$^{-3}$, is above 10\% for all models.

\begin{figure}
  \centering
  \includegraphics[width=\linewidth,angle=0]{ 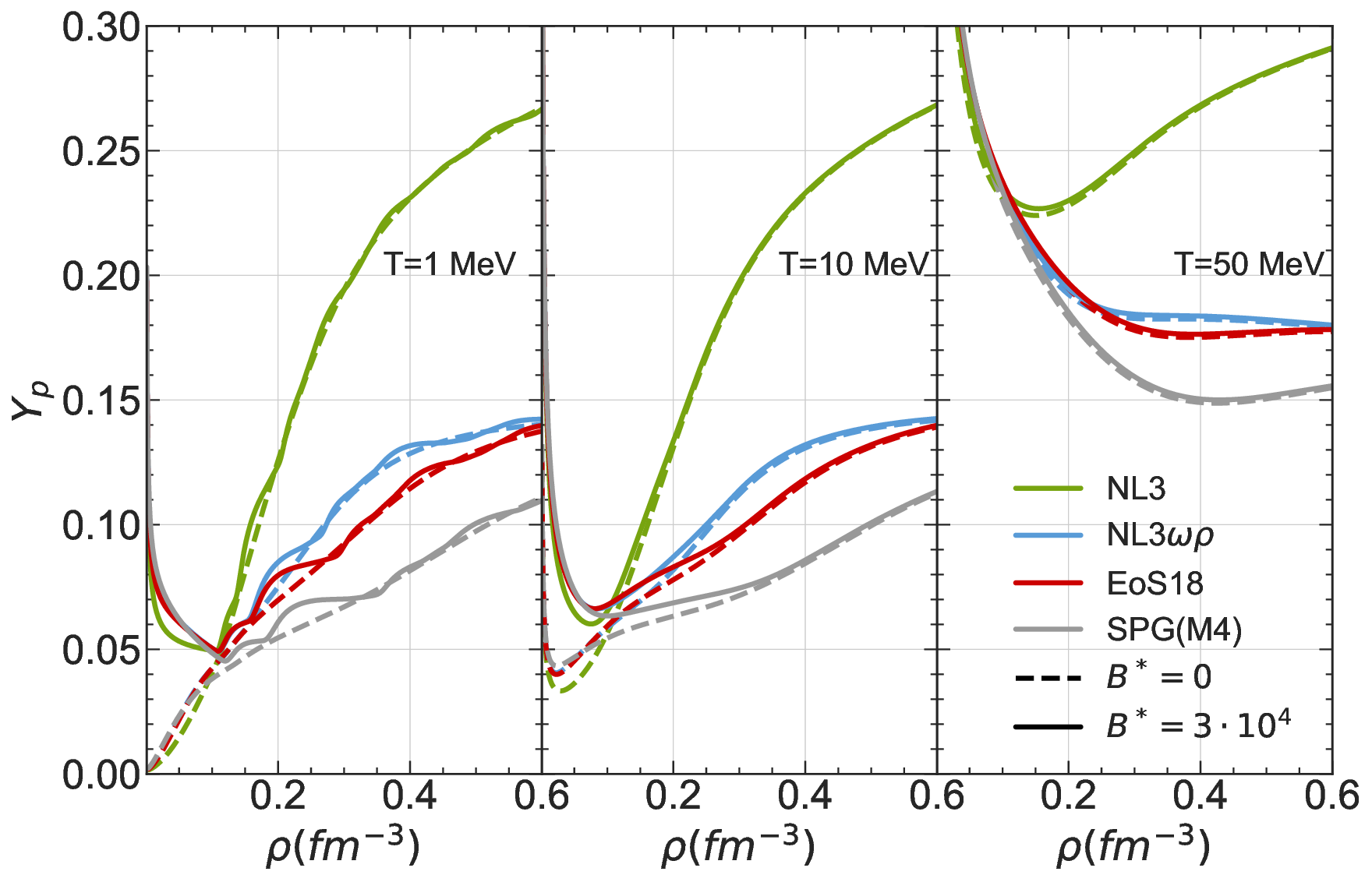} 
  \caption{Proton fraction as a function of the density for $B^*=0$ (dashed) and $B^*=3\times 10^4$ (solid). The results for $T=1 $~MeV (left), $T=10 $~MeV (center) and $T=50$~MeV (right) and for the four models are shown.}
    \label{Fig_Yp_BT}
\end{figure}

\begin{figure}
  \centering
  \includegraphics[width=\linewidth,angle=0]{ 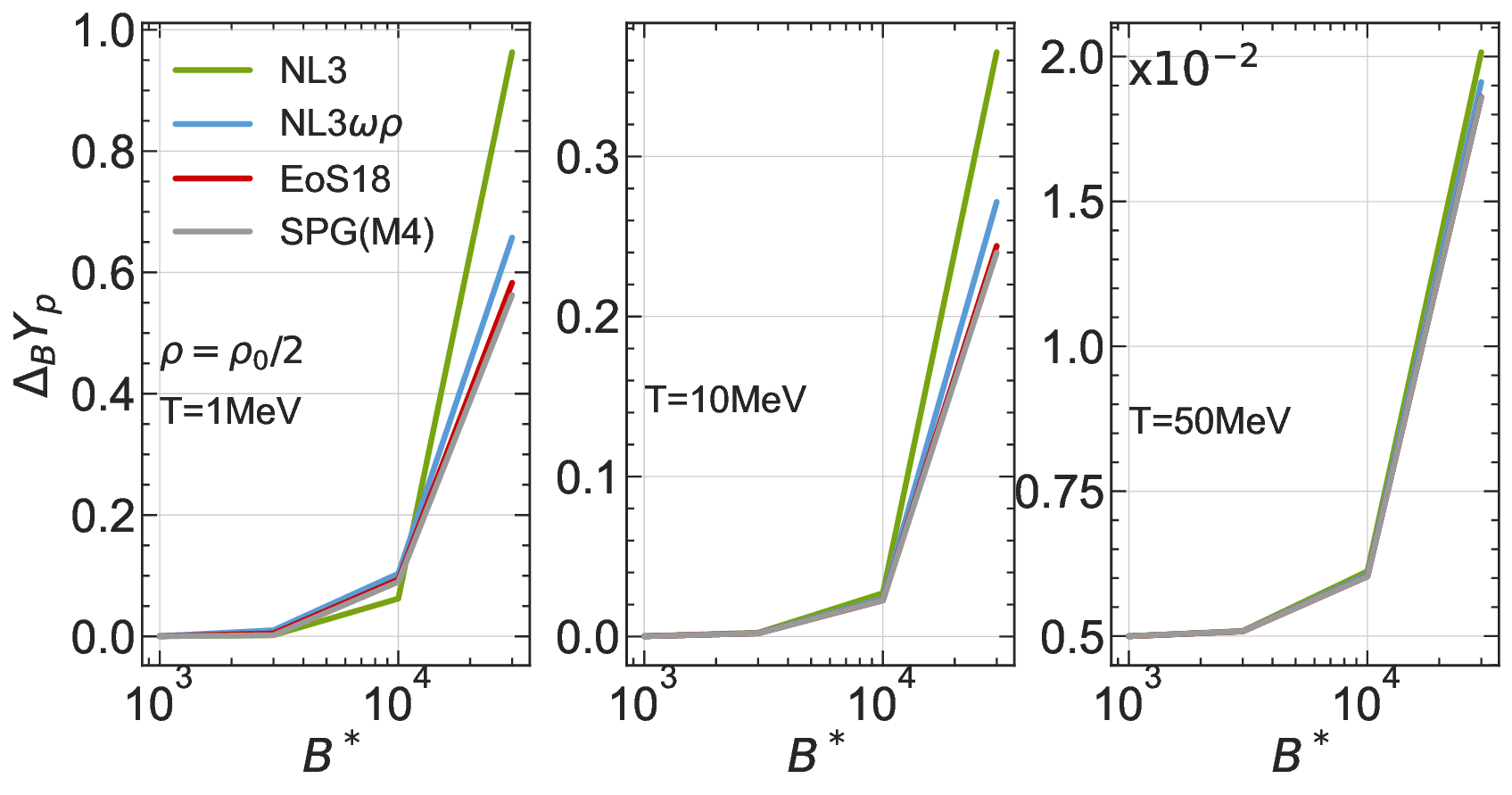} 
  \caption{Variation of the proton fraction with B (as defined in Eq.~\ref{B_var}) as a function of the magnetic field, calculated at half saturation density. The results for $T=1 $~MeV (left), $T=10 $~MeV (center) and $T=50$~MeV (right, note the difference in the scale of the $y$-axis) and for the four models are shown. }
    \label{Fig_Yp_BT_B}
\end{figure}

Fig.\ref{Fig_Yp_BT} shows the effect of the magnetic field on the composition of warm matter. We compare the proton fraction for $B^*=0$ and for a high value of the field, for three values of temperature. For $T=1$ MeV we see a clear increase in the proton fraction at low densities and the kink due to Landau quantization at higher densities. For $T=10$ MeV we still have the increase of the proton fraction at low densities, but the transitions due to Landau quantization are washed out. Finally, at very high temperatures ($T=50$ MeV) almost all effects of the magnetic field are washed out. 

The effect of the magnetic field at low densities is shown in more detail in Fig.~\ref{Fig_Yp_BT_B}, where we plot the relative variation of the proton fraction with $B$ (as defined in Eq.~\ref{B_var}) as a function of the magnetic field and calculated at $\rho=\rho_0/2$, where $\rho_0$ is the nuclear saturation density of each model. From the plot we can immediately see how the contribution of the field decreases with temperature, decreasing by almost two orders of magnitude between $T=1$ MeV and $T=50$ MeV for the highest value of the magnetic field (note the difference of scale on the $y$-axis). We also see that for the lowest temperature considered, fields below $10^{17}$ G give a variation of about 5-10\%, while for $T= 10$ MeV, these values of the field give a variation of less than 2\%, and very strong magnetic fields are needed to make a significant contribution. For $T=50$ MeV, an effect of $\sim2\%$ is only possible with magnetic fields $B\gtrsim 10^{18}$G.

In all the figures we have shown results for four different RMF models.  Although the overall behaviour of the different physical quantities is similar for the four models, we summarize some differences that should be commented on: i) The effect at low densities of the magnetic field $B$ is stronger for the model NL3 than for the other models, because this is the model with the softest symmetry energy below 0.1 fm$^{-3}$, and therefore with the smallest proton fraction at low densities. At high densities, NL3 is the model with the hardest symmetry energy and predicts the largest proton fractions; ii) at high densities, SPG(M4) has the softest symmetry energy and this is reflected in the proton fraction which is the smallest above $\sim 0.1$~fm$^{-3}$, see Fig.\ref{Fig_Yp_BT}; iii) NL3 and NL3$\omega\rho$ models have the same behaviour for symmetric nuclear matter and consequently at high densities, where the effect of isospin asymmetry is not so strong, the models show similar behaviour.  

\section{Conclusions \label{conclusions}}

In our work we study the effect of strong magnetic fields on the EoS of warm, beta-equilibrated homogeneous NS matter. We use a RMF approximation for the description of NS matter, including both models with meson-meson interaction terms and models with density dependent nucleon-meson couplings. Our results are qualitatively the same for all the models considered, showing no strong dependency on the details of the RMF models.

We start our analysis by studying the effect of magnetic fields and finite temperatures on the EoS separately. We show that the effect of temperature on both the energy per baryon and the pressure is significantly stronger than the effect of magnetic fields, with the variation of the pressure caused by the lowest temperature considered ($T= 1$ MeV) being one order of magnitude bigger than the variation caused by the highest value of the magnetic field considered ($B^*=3\times 10^4$). This shows that, when finite temperatures are taken into account, it is a safe approximation to ignore the magnetic field contribution, even for low temperatures, when calculating these two quantities.\\

We then proceed to study the effect of temperatures and magnetic fields on the fraction of protons. We show that, in both cases, the main effect is an increase in the proton fraction at low densities. This has already been shown in the case of cold magnetized matter \cite{Wang2022,Scurto_2023} and of warm non magnetized matter \cite{Tonetto2022}. We confirm these results and show that, in this case, the effect of the magnetic field remains relevant also at finite temperatures. In particular, we show that, for low temperatures ($T=1$ MeV), magnetic fields with intensity $\leq 10^{17}$ G are already enough to see a noticeable effect at low densities, while for stronger magnetic fields, it is possible to observe also the effects of the Landau quantization at higher densities. For values of the temperatures around 10 MeV only very strong magnetic fields have a significant effect (an effect of $\sim 5\%$ for $\gtrsim 5\times 10^{17}$ G and a density below half saturation density), and the effects at higher densities are almost completely negligible. Finally, at very high temperatures (T=50 MeV) effects of the magnetic field are almost washed out (an effect of 2\% requires a magnetic field $B\gtrsim 10^{18}$~G at densities below half saturation density). In this case,  it becomes a safe approximation to neglect the presence of the field even for very strong magnetic fields.

We thus conclude that the main effect of magnetic fields regards the internal composition of NSs rather than their EoS, and this effect remains relevant also at finite temperatures, if the temperatures considered are not too high.

\section{Acknowledgements}

 L.S. acknowledges the PhD grant 2021.08779.BD (FCT, Portugal)  with DOI identifier 10.54499/2021.08779.BD. Partial support from the IN2P3 Master Project NewMAC and the ANR project GWsNS, contract ANR-22-CE31-0001-01 is also acknowledged.
This work was also partially supported by national funds from FCT (Fundação para a Ciência e a Tecnologia, I.P, Portugal) under projects 
UIDB/04564/2020 and UIDP/04564/2020, with DOI identifiers 10.54499/UIDB/04564/2020 and 10.54499/UIDP/04564/2020, respectively, and the project 2022.06460.PTDC with the associated DOI identifier 10.54499/2022.06460.PTDC. H.P. acknowledges the grant 2022.03966.CEECIND (FCT, Portugal) with DOI identifier 10.54499/2022.03966.CEECIND/CP1714/CT0004.

\bibliographystyle{apsrev4-1}
\bibliography{main.bib}

\end{document}